\documentclass[draftcls,onecolumn]{IEEEtran}

\usepackage{cite}
\usepackage[pdftex]{graphicx}
\usepackage[cmex10]{amsmath}
\usepackage{amssymb}
\usepackage{amsfonts}

\usepackage{algorithm}
\usepackage{algorithmic}
\usepackage[caption=false,font=footnotesize]{subfig}

\usepackage{color}
\usepackage{array}
\usepackage{mdwmath}
\usepackage{mdwtab}
\usepackage{eqparbox}
\usepackage{lineno}

\begin{document}

\title{Big Data Analytics in Future Internet of Things}

\author{\IEEEauthorblockN{Guoru Ding, Long Wang, Qihui~Wu\\
College of Communications Engineering, \\
PLA University of Science and Technology, Nanjing 210007, China\\
email: dingguoru@gmail.com; wanglong127@gmail.com; wqhtxdk@aliyun.com}

\thanks{This work is supported by the National Natural Science Foundation of China (Grant No. 61172062, 60932002, No. 6131160), the National Basic Research Program of China (Grant No. 2009CB320400), and in part by Jiangsu Province Natural Science Foundation (Grant No. BK2011116).}}
\maketitle

\begin{abstract}
Current research on Internet of Things (IoT) mainly focuses on how to enable general objects to see, hear, and smell the physical world for themselves, and make them connected to share the observations. In this paper, we argue that only connected is not enough, beyond that, general objects should have the capability to learn, think, and understand both the physical world by themselves. On the other hand, the future IoT will be highly populated by large numbers of heterogeneous networked embedded devices, which are generating massive or big data in an explosive fashion. Although there is a consensus among almost everyone on the great importance of big data analytics in IoT, to date, limited results, especially the mathematical foundations, are obtained. These practical needs impels us to propose a systematic tutorial on the development of effective algorithms for big data analytics in future IoT, which are grouped into four classes: 1) heterogeneous data processing, 2) nonlinear data processing, 3) high-dimensional data processing, and 4) distributed and parallel data processing. We envision that the presented research is offered as a mere baby step in a potentially fruitful research direction. We hope that this article, with interdisciplinary perspectives, will stimulate more interests in research and development of practical and effective algorithms for specific IoT applications, to enable smart resource allocation, automatic network operation, and intelligent service provisioning.
\end{abstract}

\IEEEpeerreviewmaketitle


\section{Introduction}

The \emph{Internet of Things (IoT)}, firstly coined by Kevin Ashton as the title of a presentation in 1999~\cite{Ashton-1999}, is a technological revolution that is bringing us into a new ubiquitous connectivity, computing, and communication era. The development of IoT depends on dynamic technical innovations in a number of fields, from wireless sensors to nanotechnology~\cite{ITU-2005}. For these ground-breaking innovations to grow from ideas to specific products or applications, in the past decade, we have witnessed worldwide efforts from academic community, service providers, network operators, and standard development organizations, etc (see, e.g., the recent comprehensive surveys in~\cite{ASurvey,Context-Survey,Standardized-survey}). Generally, current research on IoT mainly focuses on how to enable general objects to \emph{see, hear, and smell} the physical world for themselves, and make them connected to share the observations. In this paper, we argue that only connected is not enough, beyond that, general objects in future IoT should have the capability to \emph{learn, think, and understand} the physical world by themselves.

Specifically, the future IoT will be highly populated by large numbers of heterogeneous networked embedded devices, which are generating massive or big data in an explosive fashion. The big data we collect may not have any value unless we analyze, interpret, understand, and properly exploit it. Although there is a consensus among almost everyone on the great importance of big data analytics in IoT, to date, limited results, especially the mathematical foundations, are obtained.

\begin{figure}[!h]
\centering
\includegraphics[width=4.5in]{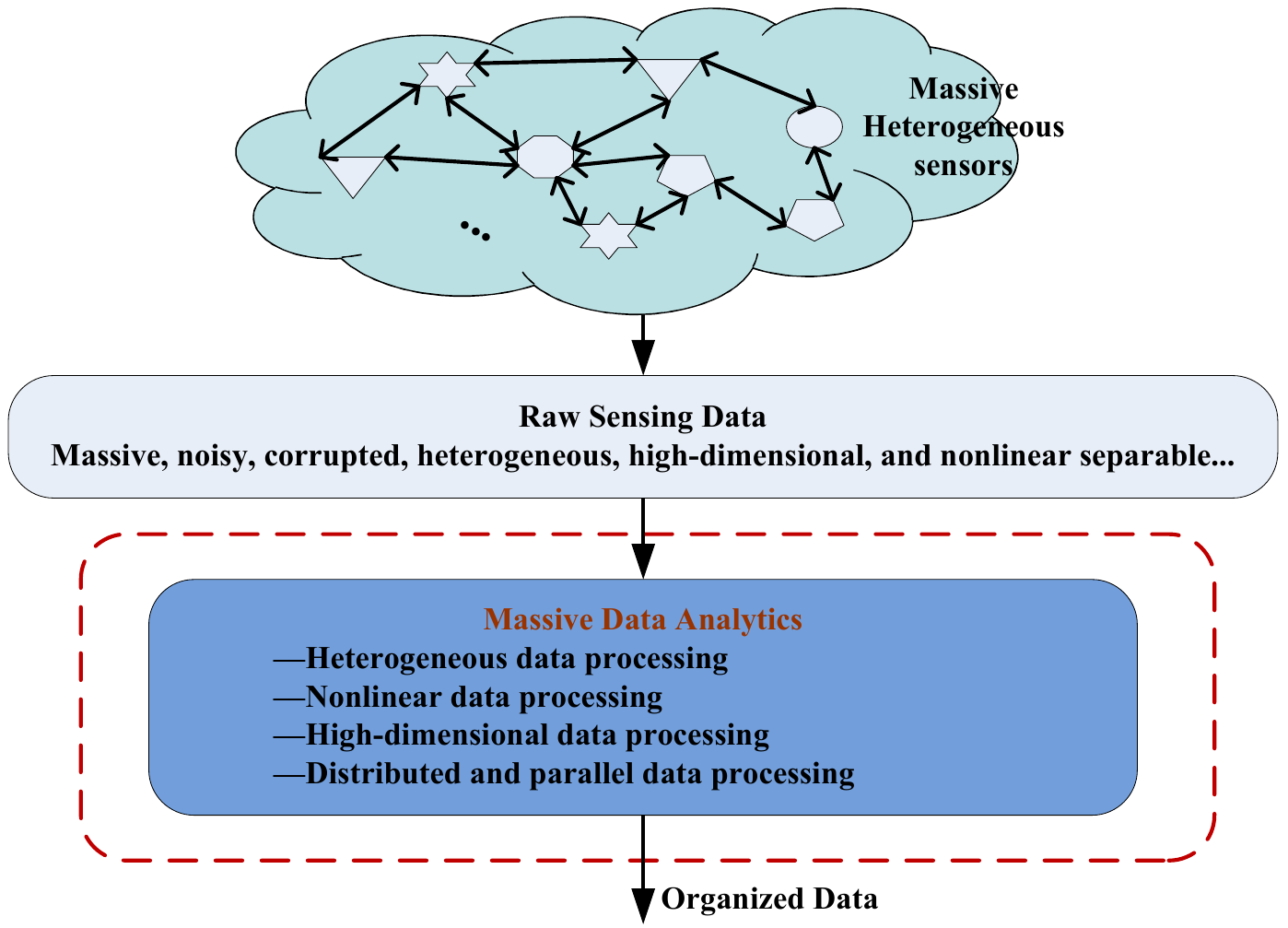}
\caption{The research framework of massive data analytics in IoT.}
\label{Visio-framework-massive}
\end{figure}

As shown in Fig.~\ref{Visio-framework-massive}, in this paper we propose a systematic tutorial on the development of effective algorithms for big data analytics in future IoT, which are grouped into four classes: 1) heterogeneous data processing, 2) nonlinear data processing, 3) high-dimensional data processing, and 4) distributed and parallel data processing.

\section{Application Scenario}
\begin{figure}[!h]
\centering
\includegraphics[width=3in]{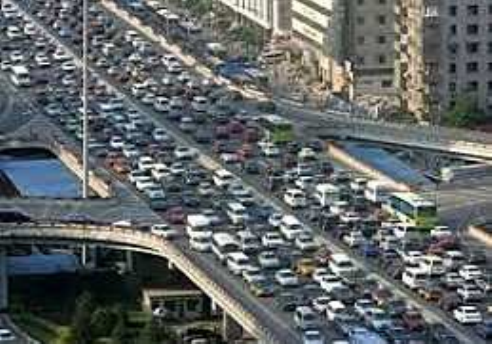}
\caption{A motivational and illustrative application scenario of IoT.}
\label{scenario}
\end{figure}

Before gonging deep into the technical details, let's first share one interesting application scenario that will i) facilitate the understanding of the following sections and ii) probably come into our daily life in future:

As shown in Fig. \ref{scenario}, living in a modern city, traffic jams harass many of us. With potential traffic jams into consideration, every time when the source and the destination is clear, it is generally not easy for a driver to decide what the quickest route should be, especially when the driver is fresh to the city. Among many others, the following scheme may be welcome and useful for drivers: \emph{Suppose that there are a city of crowdsourcers, such as pre-deployed cameras, vehicles, drivers, and/or passengers, intermittently observe the traffic flow nearby and contribute their observations to a data center. The data center effectively fuses the crowdsourced observations to generate real-time traffic situation map and/or statistical traffic database. Then, every time when a driver tells his/her car the destination, the car will automatically query the data center, deeply analyze the accessed traffic situation information from the data center and meanwhile other cars/drivers' potential decisions, and intelligently selects the quickest route or a few top quickest routes for its driver.}

\section{Heterogeneous Data Processing}

In practical IoT applications, the massive data are generally collected from heterogeneous sensors (e.g., cameras, vehicles, drivers, and passengers), which in turn may provide heterogeneous sensing data (e.g., text, video, and voice). Heterogeneous data processing (e.g., fusion, classification) brings unique challenges and also offers several advantages and new possibilities for system improvement.

Mathematically, random variables that characterize the data from heterogeneous sensors may follow disparate probability distributions. Denote $z_n$ as the data from the $n$-th sensor and ${\bf{Z}}:=\{z_n\}_{n=1}^N$ as the heterogeneous data set, the marginals $\{z_n\}_{n=1}^N$ are generally non-identically or heterogeneously distributed. In many IoT applications, problems are often modeled as multi-sensor data fusion, distribution estimation or distributed detection. In these cases, joint probability density function (pdf) $f({\bf{Z}})$ of the heterogeneous data set ${\bf{Z}}$ is needed to obtain from the marginal pdfs $\{f(z_n)\}_{n=1}^N$.


For mathematical tractability, one often chooses to assume simple models such as the product model or multivariate Gaussian model, which lead to suboptimal solutions~\cite{Correlation}. Here we recommend another approach, based on copula theory, to tackle heterogeneous data processing in IoT. In copula theory, it is the copulas function that couples multivariate joint distributions to their marginal distribution functions, mainly thanks to the following theorem:

\emph{Sklar' Theorem~\cite{copula}:} Let $F$ be an $N$-dimensional cumulative distribution function (cdf) with continuous marginal cdfs $F_1,F_2,...,F_N$. Then there exists a unique copulas function $C$ such that for all $z_1,z_2,...,z_N$ in $[-\infty,+\infty]$
\begin{align}
\label{eq:Copula}
F(z_1,z_2,...,z_N)=C\big(F_1(z_1),F_2(z_2),...,F_N(z_N)\big).
\end{align}

The joint pdf can now be obtained by taking the $N$-order derivative of (\ref{eq:Copula})
\begin{align}
\label{eq:Copula2}
f(z_1,z_2,...,z_N)&=\frac{\partial^N}{\partial_{z_1}\partial_{z_2}...\partial_{z_N}}C\big(F_1(z_1),F_2(z_2),...,F_N(z_N)\big)\nonumber\\
&=f_p(z_1,z_2,...,z_N)c\big(F_1(z_1),F_2(z_2),...,F_N(z_N)\big),
\end{align}
where $f_p(z_1,z_2,...,z_N)$ denotes the product of the marginal pdfs $\{f(z_n)\}_{n=1}^N$ and $c(\cdot)$ is the copula density weights the product distribution appropriately to incorporate dependence between the random variables. The topic on the design or selection of proper copula functions is well summarized in~\cite{copula2}.

\section{Nonlinear Data Processing}

In IoT applications such as multi-sensor data fusion, the optimal fusion rule can be derived from the multivariate joint distributions obtained in (\ref{eq:Copula2}). However, it is generally mathematically intractable since the optimal rule generally involves nonlinear operations~\cite{Data-Fusion}. Therefore, linear data processing methods dominate the research and development, mainly for their simplicity. However, linear methods are often oversimplified to deviate the optimality.

In many practical applications, nonlinear data processing significantly outperforms their linear counterparts. Kernel-based learning (KBL) provides an elegant mathematical means to construct powerful nonlinear variants of most well-known statistical linear techniques, which has recently become prevalent in many engineering applications~\cite{SPMag2013}.

Briefly, in KBL theory, data ${\bf{x}}$ in the input space $\mathcal{X}$ is projected onto a higher dimensional feature
space $\mathcal{F}$ via a \emph{nonlinear mapping} $\Phi$ as follows:
\begin{align}
\label{eq:nonlinear-map}
\Phi:~\mathcal{X} ~ \to ~\mathcal{F},~~~~{\bf{x}}~ \mapsto ~ \Phi({\bf{x}}).
\end{align}

\begin{figure}[!t]
\centering
\includegraphics[width=3.5in]{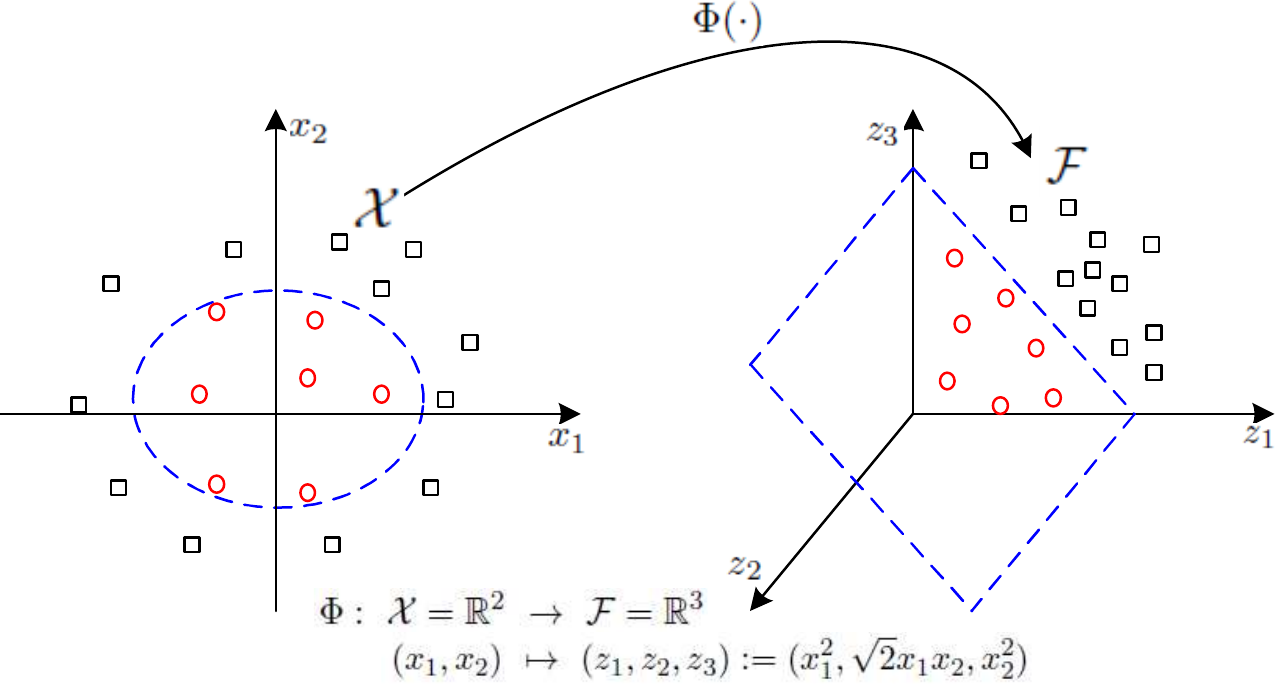}
\caption{An introductory binary classification example~\cite{SPMag2013}. By mapping data ${\bf{x}}=(x_1,x_2)$ in 2-D input space $\mathcal{X}=\mathbb{R}^2$ (left) via nonlinear mapping $\Phi(\cdot)$ onto a 3-D feature space $\mathcal{F}=\mathbb{R}^3$ (right), the data become linearly separable.}
\label{fig_mapping}
\end{figure}


For a given problem, one now works with the mapped data $\Phi({\bf{x}}) \in \mathcal{F}$ instead of ${\bf{x}} \in \mathcal{X}$. The data in the input space can be projected onto different feature spaces with different mappings. The diversity of feature spaces provides us more choices to gain better performance. Actually, without knowing the mapping $\Phi$ explicitly, one only needs to replace the inner product operator of a linear technique with an appropriate~\emph{kernel} ${\rm{k}}$ (i.e., a positive semi-definite symmetric function),
\begin{align}
\label{eq:kernel}
{\rm{k}}({\bf{x}}_i,{\bf{x}}_j) := \langle\Phi({\bf{x}}_i), \Phi({\bf{x}}_j)\rangle_{\mathcal{F}}, ~~\forall {\bf{x}}_i,{\bf{x}}_j \in\mathcal{X}.
\end{align}

The most widely used kernels can be divided into two categories: projective kernels (functions of inner product, e.g., polynomial kernels) and radial kernels (functions of distance, e.g., Gaussian kernels)~\cite{SPMag2013}.

\section{High-Dimensional Data Processing}

In IoT, massive or big data always accompanies high-dimensionality. For example, images and videos observed by cameras in many IoT applications are generally very high-dimensional data, where the dimensionality of each observation is comparable to or even larger than the number of observations. Moreover, in kernel-based learning methods discussed above, the kernel function nonlinearly maps the data in the original space into a higher dimensional feature space, which transforms virtually every dataset to a high-dimensional one.

Mathematically, we can represent the massive data in a compact matrix form. Many practical applications have experimentally demonstrated the intrinsic low-rank property of the high-dimensional data matrix, such as the traffic matrix in large scale networks~\cite{Tracking-Network} and image frame matrix in video surveillance~\cite{RPCA}, which is mainly due to common temporal patterns across columns or rows, and periodic behavior across time, etc.

Low-rank matrix plays a central role in large-scale data analysis and dimensionality reduction. In the following, we provide a brief tutorial on using low-rank matrix recovery and/or completion\footnote{Matrix completion aims to recover the missing entries of a matrix, given limited number of known entries, while matrix recovery aims to recover the matrix with corrupted entries.} algorithms for high-dimensional data processing, from simple to complex.

\subsubsection{Low-rank matrix recovery with dense noise and sparse anomalies}

Suppose we are given a large sensing data matrix ${\bf{Y}}$, and know that it may be decomposed as
\begin{align}
{\bf{Y}} = {\bf{X}}+{\bf{V}},
\end{align}
where ${\bf{X}}$ has low-rank, and ${\bf{V}}$ is a perturbation/noise matrix with entry-wise non-zeros. We do not know the low-dimensional column and row space of ${\bf{X}}$, not even their dimensions. To stably recover the matrix ${\bf{X}}$ from the sensing data matrix ${\bf{Y}}$, the problem of interest can be formulated as classical principal component analysis (PCA):
\begin{align}
\label{eq:Opt1}
\min_{\{{\bf{X}}\}}~||{\bf{X}}||_*~~~{\text{subject~to}}~||{\bf{Y}}-{\bf{X}}||_F \le \varepsilon,
\end{align}
where $\varepsilon$ is a noise related parameter, $||\cdot||_*$ and $||\cdot||_F$ stands for the nuclear norm (i.e., the sum of the singular values) and the Frobenious norm of a matrix.

Furthermore, if there are also some abnormal data ${\bf{A}}$ injected into the sensing data matrix ${\bf{Y}}$, we have
\begin{align}
{\bf{Y}} = {\bf{X}}+{\bf{V}}+{\bf{A}},
\end{align}
where ${\bf{A}}$ has sparse non-zero entries, which can be of arbitrary magnitude. In this case, we do not know the low-dimensional column and row space of ${\bf{X}}$, not know the locations of the nonzero entries of ${\bf{A}}$, and not even know how many there are. To accurately and efficiently recover the low-rank data matrix ${\bf{X}}$ and sparse component ${\bf{A}}$, the problem of interest can be formulated as the following tractable convex optimization:
\begin{align}
\label{eq:opt2}
\min_{\{{\bf{X}},{\bf{A}}\}}~||{\bf{X}}||_{*}+\lambda ||{\bf{A}}||_{1}~~~{\text{subject~to}}~||{\bf{Y}}-{\bf{X}}-{\bf{A}}||_F \le \varepsilon,
\end{align}
where $\lambda$ is a positive rank-sparsity controlling parameter, and $||\cdot||_1$ stands for the $l_1$-norm (i.e., the number of nonzero entries) of a matrix.

\subsubsection{Joint matrix completion and matrix recovery}
In practical IoT applications, it is typically difficult to acquire all entries of the sensing data matrix {\bf{Y}}, mainly due to i) transmission loss of the sensing data from the sensors to the data center, and ii) lack of incentives for the crowdsourcers to contribute all their sensing data.

In this case, the sensing data matrix $\widetilde{\bf{Y}}$ is made up of \emph{noisy}, \emph{corrupted}, and \emph{incomplete} observations,
\begin{align}
\widetilde{\bf{Y}}:={\mathcal{P}}_{\Omega}({\bf{Y}})={\mathcal{P}}_{\Omega}({\bf{X}}+{\bf{A}}+{\bf{V}}),
\end{align}
where $\Omega \subseteq [M]\times[N]$ is the set of indices of the acquired entries, and $\mathcal{P}_\Omega$ is the orthogonal projection onto the linear subspace of matrices supported on $\Omega$, i.e., if $(m,n) \in  {\Omega}$, ${\mathcal{P}}_{\Omega}({\bf{Y}}) ={y}_{m,n}$; otherwise, ${\mathcal{P}}_{\Omega}({\bf{Y}}) =0$. To stably recover the low-rank and sparse components ${\bf{X}}$ and ${\bf{A}}$, the problem can be further formulated as
\begin{align}
\label{eq:opt-com1}
\min_{\{{\bf{X}},{\bf{A}}\}}~||{\bf{X}}||_{*}+\lambda ||{\bf{A}}||_{1}~~~{\text{subject~to}}~||{\mathcal{P}}_{\Omega}({\bf{Y}})-{\mathcal{P}}_{\Omega}({\bf{X}}+{\bf{A}}+{\bf{V}})||_F \le \varepsilon.
\end{align}

All the problems formulated in (\ref{eq:Opt1}), (\ref{eq:opt2}), and (\ref{eq:opt-com1}) fall into the scope of convex optimization, efficient algorithms can be developed based on the results in~\cite{ALM}.

\section{Parallel and Distributed Data Processing}

So far, all the data processing methods introduced above are in essence centralized and suitable to be implemented at a data center. However, in many practical IoT applications, where the objects in the networks are organized in an \emph{ad hoc} or decentralized manner, centralized data processing will be inefficient or even impossible because of single-node failure, limited scalability, and huge exchange overhead, etc. Now, one natural question comes into being: Is there any way to disassemble massive data into groups of small data, and transfer centralized data processing into decentralized processing among locally interconnected agents, at the price of affordable performance loss?

In this subsection, we argue that alternating direction method of multipliers (ADMM)~\cite{ADMM,BOOK-DB} serves as a promising theoretical framework to accomplish parallel and distributed data processing. Suppose a very simple case with a IoT consisting of $N$ interconnected smart objects. They have a common objective as follows
\begin{align}
\label{ADMM1}
\min_{\bf{x}} f({\bf{x}})=\sum_{i=1}^N f_i({\bf{x}}),
\end{align}
where ${\bf{x}}$ is an unknown global variable and $f_i$ refers to the term with respect to the $i$-th smart object. By introducing local variables $\{{\bf{x}}_i\in {\bf{R}}^n\}_{i=1}^N$ and a common global variable ${\bf{z}}$, the problem in (\ref{ADMM1}) can be rewritten as
\begin{align}
\label{ADMM2}
\min_{\{{\bf{x}}_1,...,{\bf{x}}_N,{\bf{z}}\}} \sum_{i=1}^N f_i({\bf{x}}_i)~~~{\text{subject~to}}~{\bf{x}}_i={\bf{z}}, ~~i = 1,...,N.
\end{align}

This is called the global consensus problem, since the constraint is that all the local variables should agree, i.e., be equal. The augmented Lagrangian of problem (\ref{ADMM2}) can be further written as
\begin{align}
\label{ADMM3}
L_{\mu} ({\bf{x}}_1,...,{\bf{x}}_N,{\bf{z}},{\bf{y}}) = \sum_{i=1}^N \big(f_i({\bf{x}}_i)+ {\bf{y}}_i^T({\bf{x}}_i-{\bf{z}})+\frac{\mu}{2}||{\bf{x}}_i-{\bf{z}}||_F^2  \big).
\end{align}

The resulting ADMM algorithm directly from (\ref{ADMM3}) is the following:
\begin{align}
&{\bf{x}}_i^{k+1} := \text{argmin}_{{\bf{x}}_i} \big( f_i({\bf{x}}_i)+ {\bf{y}}_i^{kT}({\bf{x}}_i-{\bf{z}}^k)+\frac{\mu}{2}||{\bf{x}}_i-{\bf{z}}^k||_F^2\big)\\
&{\bf{z}}^{k+1} := \frac{1}{N} \sum_{i=1}^N \big( {\bf{x}}_i^{k+1} +1/\mu {\bf{y}}_i^{k} \big)\\
&{\bf{y}}_i^{k+1}:={\bf{y}}_i^{k}+\mu( {\bf{x}}_i^{k+1}- {\bf{z}}^{k+1}).
\end{align}

The first and last steps are carried out independently at each smart object, while the second step is performed at a fusion center. Actually, when the smart objects are multi-hop connected, the second step can be replaced by
\begin{align}
\label{ADMM4}
{\bf{z}}_i^{k+1} := \frac{1}{|\mathcal{N}_i|} \sum_{i\in\mathcal{N}_i} \big( {\bf{x}}_i^{k+1} +1/\mu {\bf{y}}_i^{k} \big)
\end{align}
where $\mathcal{N}_i$ denotes the one-hop neighbor set of the $i$-th object and $|\cdot|$ is the cardinality of a set. Eq. (\ref{ADMM4}) means that the second step can also be carried out at each smart object by fusing the local data from one-hop neighbors.

This is a very intuitive algorithm to show the basic principle of ADMM. More advanced algorithms can be developed according to the theoretical results in~\cite{ADMM,BOOK-DB}.

\section{Conclusions and Discussions}
In this paper, we have provided a high-level tutorial on massive data analytics in terms of heterogeneous, nonlinear, high-dimensional, and distributed and parallel data processing, respectively. Actually, in practical IoT applications, the obtained massive sensing data can be of mixed characteristics, which is much more challenging. Moreover, the development of practical and effective algorithms for specific IoT applications are also urgently needed. We envision that the presented research is offered as a mere baby step in a potentially fruitful research direction. We hope that this article, with interdisciplinary perspectives, will stimulate more interests in research and development of practical and effective algorithms for big data analytics in IoT, to enable smart resource allocation, automatic network operation, and intelligent service provisioning.

\end{document}